\documentclass[conference, a4paper]{IEEEtran}
\usepackage{cite}
\ifCLASSINFOpdf
  \usepackage[pdftex]{graphicx}
\else
  \usepackage[dvips]{graphicx}
\fi
\usepackage[cmex10]{amsmath}
\usepackage{array}
\usepackage{theorem}
\usepackage[caption=false,font=footnotesize]{subfig}
\usepackage{fixltx2e}
\usepackage{url}

\begin{document}

\title{Non-Systematic Codes for \\ Physical Layer Security}

\author{\IEEEauthorblockN{Marco Baldi, Marco Bianchi, Franco Chiaraluce,}
\IEEEauthorblockA{DIBET, Polytechnic University of Marche,\\
Ancona, Italy\\
Email: \{m.baldi, m.bianchi, f.chiaraluce\}@univpm.it}}

\maketitle

\begin{abstract}
This paper is a first study on the topic of achieving physical layer security
by exploiting non-systematic channel codes.
The chance of implementing transmission security at the physical layer is known since 
many years in information theory, but it is now gaining an increasing interest
due to its many possible applications.
It has been shown that channel coding techniques can be effectively exploited for
designing physical layer security schemes, able to ensure that an unauthorized receiver,
experiencing a channel different from that of the the authorized receiver, is not able to gather
any information.
Recently, it has been proposed to exploit puncturing techniques in order to reduce
the security gap between the authorized and unauthorized channels.
In this paper, we show that the same target can also be achieved by using non-systematic
codes, able to scramble information bits within the transmitted codeword.
\end{abstract}

\section{Introduction}
Despite nowadays transmission security is often implemented at higher layers, the idea 
of achieving it at physical layer has been the inspiring basis for the development 
of the modern theory of cryptography.
When security is implemented at physical layer, all receivers
share the same (complete) knowledge of the transmission technique, without the need
of any form of secret sharing.
The channel is responsible for differentiation among users, and security is only based
on the differences among the channels experienced by authorized and unauthorized users.

\begin{figure}[t]
\begin{centering}
\includegraphics[width=83mm,keepaspectratio]{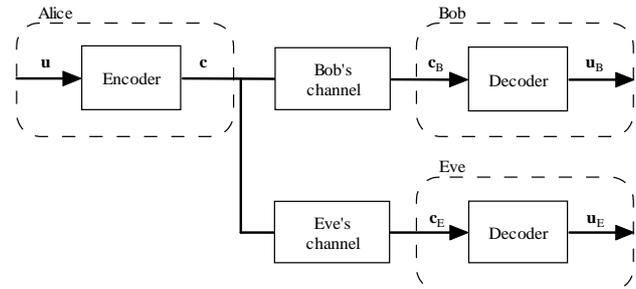}
\caption{Block scheme of a wire-tap channel. \label{fig:WireTap}}
\par\end{centering}
\end{figure}

A very simple model that is well suited to represent physical layer security schemes is the
\textit{wire-tap channel}, first introduced by Wyner in 1975 \cite{Wyner1975}.
In the wire-tap channel model, a transmitter (Alice) sends information to the
legitimate receiver (Bob), but this is also received by the eavesdropper (Eve).
Alice can adopt whatever randomization, encoding and modulation scheme before transmitting
her message, and both Bob and Eve are perfectly aware of the transmission technique she uses; 
so, at least in principle, they are both able to recover the plaintext message ($\mathbf{u}$) 
from the ciphertext ($\mathbf{c}$).
However, the channel that separates Alice from Bob is generally different from that between Alice
and Eve. For this reason, the ciphertext received by Bob ($\mathbf{c}_\mathrm{B}$) is different
from that gathered by Eve ($\mathbf{c}_\mathrm{E}$).
So, after inverting the encoding map, the message obtained by Bob ($\mathbf{u}_\mathrm{B}$)
can differ from that recovered by Eve ($\mathbf{u}_\mathrm{E}$).
A block scheme of the wire-tap channel is reported in Fig. \ref{fig:WireTap}.

Based on these assumptions, physical layer security on the wire-tap channel is achieved
when Bob is able to exactly reconstruct the plaintext message, i.e., $\mathbf{u}_\mathrm{B} = \mathbf{u}$, 
with very high probability,
whereas the message recovered by Eve has no form of correlation with $\mathbf{u}$.
An important parameter for the wire-tap channel is the secrecy capacity, defined as the
highest transmission rate at which Eve is unable to decode any information.

A lot of literature is devoted to studying the wire-tap channel capacity and how to achieve it.
For a detailed discussion on the wire-tap channel and the secrecy capacity, we refer the 
interested reader to \cite{Maurer1993, Thangaraj2007} and the references therein.
The literature confirms that an increasing attention has been devoted to the application of classic and modern coding techniques on the wire-tap channel \cite{Thangaraj2007, Merhav2008, 
Klinc2009, Kwak2009, Harrison2009}.

In this study, we are interested in the Additive White Gaussian Noise (AWGN) wire-tap model, where the secrecy capacity
equals the difference between the two channel capacities \cite{Leung-Yan-Cheong1978}.
So, in order to achieve transmission security, Bob's channel must have higher 
signal-to-noise ratio (SNR) than Eve's channel. Alternatively, a feedback channel 
between Alice and Bob (also accessible to Eve) is needed \cite{Lai2008}, but such case
is not considered in this paper.

In the context defined above, an important parameter is the \textit{security gap}, that expresses
the quality difference between Bob's and Eve's channels required
to achieve a sufficient level of physical layer security.
An important target is to keep the security gap as small as possible, in such a way
as to achieve physical layer security even with a small degradation of Eve's channel
with respect to Bob's one.
Some recent literature has been devoted to the study of the way such reduction can be
achieved by exploiting techniques from coding theory \cite{Klinc2009}.

In this paper, we investigate how the security gap can be reduced by exploiting
non-systematic transmission, in which the information bits are not in clear within
the transmitted codewords, but are scrambled during encoding. In particular we show that such solution, suitably combined with error correcting codes, can outperform other schemes, recently proposed, based on punctured codes, as it provides smaller secrecy gaps for a given set of parameters. Both the cases of hard-decoded classic block codes (e.g., BCH codes) and modern soft-decoded block codes (e.g., LDPC codes) are considered and reveal to be equally effective.

The paper is organized as follows. In Section \ref{sec:two} we introduce the notation. In Section \ref{sec:three} we describe the coding scheme and the role of scrambling. 
In Section \ref{sec:four} the analysis is extended to the case of non-systematic LDPC codes. In Section \ref{sec:five} we compare the secrecy gap performance of the various solutions considered. Finally, Section \ref{sec:six} concludes the paper.

\section{Notation and Related Work}
\label{sec:two}

We consider an AWGN wire-tap channel model in which Alice sends a secret message 
in the form of a $1 \times k$ binary vector $\mathbf{u}$.
Before transmission, the secret message is encoded by Alice into a $1 \times n$ 
binary word $\mathbf{c}$, with $n \geq k$, that is then transmitted over the 
channel.
The \textit{secrecy rate} $R_s$ is defined as the ratio between the secret message
length and the transmitted word length.
So, in the case we consider, the secrecy rate coincides with the transmission rate:

\begin{equation}
R_s = R = \frac{k}{n}.
\end{equation}

More in general, it should be $R_s \leq R$, since part of the transmitted information
bits could be non-secret. The special case $R_s = R$ is considered here for the sake of simplicity.

The transmitted word is received by Bob and Eve through two different channels.
We denote by $\mathbf{c}_\mathrm{B}$ the word received by Bob and by $\mathbf{c}_\mathrm{E}$
the word received by Eve, respectively. Bob's and Eve's channels are corrupted by AWGN
with different SNR: $\left. \frac{E_b}{N_0} \right|_\mathrm{B}$ is
Bob's channel energy per bit to noise power spectral density ratio, whereas
$\left. \frac{E_b}{N_0} \right|_\mathrm{E}$ is the same parameter for Eve's channel.
Similarly, $P_e^\mathrm{B}$ is Bob's bit error rate and $P_e^\mathrm{E}$ is Eve's
one.

Security at physical layer is achieved when Bob has bit error rate lower than
a given threshold, $P_e^\mathrm{B} \leq \overline{P_e^\mathrm{B}}$, while Eve's bit 
error rate is greater than another threshold (next to $0.5$), 
$P_e^\mathrm{E} \geq \overline{P_e^\mathrm{E}}$.
Starting from the curve of bit error rate as a function of the signal-to-noise ratio
for the transmission technique adopted, these two values 
can be expressed in terms of $\frac{E_b}{N_0}$, and
the security gap $S_g$ is easily obtained as follows:

\begin{equation}
\left\{ 
\begin{array}{l}
\overline{P_e^\mathrm{B}} = f\left(\left. \overline{\frac{E_b}{N_0}} \right|_\mathrm{B}\right), \\
\overline{P_e^\mathrm{E}} = f\left(\left. \overline{\frac{E_b}{N_0}} \right|_\mathrm{E}\right), \\
S_g = \left. \overline{\frac{E_b}{N_0}} \right|_\mathrm{B} - \left. \overline{\frac{E_b}{N_0}} \right|_\mathrm{E}.\\
\end{array} 
\right.
\end{equation}

Several works have been devoted to the study of what transmission techniques are best
suited to reduce the security gap.
In particular, in \cite{Klinc2009}, the authors propose the usage of punctured codes,
by associating the secret bits to punctured bits.
They consider punctured LDPC codes and prove that such technique, 
for a fixed secrecy rate, is able to guarantee a considerable reduction in the security gap 
with respect to non-punctured (systematic) transmission.

In this paper, we consider an alternative solution, based on non-systematic coding.
As we will show in the following sections, non-systematic coding is also able to achieve
a strong reduction in the security gap, that becomes comparable (and even better) than
that obtained through puncturing.

\section{Physical Layer Security through Non-Systematic Codes}
\label{sec:three}

In the scheme we consider, Alice implements the encoding map as follows:

\begin{equation}
\mathbf{c} = \mathbf{u \cdot S \cdot G},
\end{equation}
where $\mathbf{G}$ is the $k \times n$ generator matrix of an $(n, k)$-linear block code in systematic
form, and $\mathbf{S}$ is a non-singular $k \times k$ binary scrambling matrix.
Due to its systematic character, $\mathbf{G}$ can also be written as 
$\mathbf{G} = \left[ \mathbf{I} | \mathbf{C} \right]$, where $\mathbf{I}$ is a $k \times k$
identity matrix and $\mathbf{C}$ is a $k \times \left(n-k\right)$ matrix representing
the parity-check constraints.
This settings resembles that of the McEliece cryptosystem \cite{McEliece1978}, where, in
addition, the encoded word is also permuted.

Based on these assumptions, the encoded word can also been written as
$\mathbf{c} = \left[\mathbf{u \cdot S} | \mathbf{u \cdot S \cdot C}\right] = \left[\mathbf{c}_l | \mathbf{c}_r \right]$,
where $\mathbf{c}_l$ is the vector containing the first $k$ bits of $\mathbf{c}$, while $\mathbf{c}_r$
collects its last $r$ bits.
Both Bob's and Eve's channels introduce errors. However, as mentioned, $\left. \frac{E_b}{N_0} \right|_\mathrm{B}$ must be large enough to ensure that, with very high probability, Bob's decoder is able to correct all errors, thus delivering, after descrambling, $\mathbf{u}_\mathrm{B} = \mathbf{u} = \mathbf{c}_l \cdot \mathbf{S}^{-1}$. On the contrary, $\left. \frac{E_b}{N_0} \right|_\mathrm{E}$ must be small enough to ensure that, after decoding, the codeword obtained by Eve is still affected by an error vector $\mathbf{e}$. 
So, in this case, at the output of the descrambler, Eve has:

\begin{equation}
\mathbf{u}_\mathrm{E} = \mathbf{u} + \mathbf{e}_l \cdot \mathbf{S}^{-1},
\label{eq:ErrorSpreading}
\end{equation}
where $\mathbf{e}_l$ is the left part of the error vector $\mathbf{e} = \left[\mathbf{e}_l | \mathbf{e}_r \right]$.
From \eqref{eq:ErrorSpreading} we notice that descrambling
has the effect of spreading the residual errors
after decoding.

For the goals of the present paper, it is preliminarily useful to obtain, in analytical terms, an estimate of the bit error rate 
($P_e$) and frame error rate ($P_f$) for Bob and Eve in absence or in presence of
scrambling.
For such purpose, we first refer to two explicative cases, namely, unitary rate coding and
\textit{t}-error correcting coding.
In the next section, we will provide further results, based on numerical simulations, in which 
we consider LDPC coding and we will compare the proposed approach with that based on puncturing.

\subsection{Unitary Rate Coding}

We can consider the case of unitary rate coding by imposing $k = n$ and $\mathbf{G}$
coincident with a $k \times k$ identity matrix $\mathbf{I}_k$.
If we also assume $\mathbf{S} = \mathbf{I}_k$, the unitary code is systematic, 
and the system reduces itself to a framed uncoded transmission.
Focusing attention on the case of Binary Phase Shift Keying (BPSK), the bit and frame error
probabilities are given by:

\begin{equation}
\left\{ 
\begin{array}{l}
P_e = \frac{1}{2}\mathrm{erfc}\left(\sqrt{\frac{E_b}{N_0}}\right), \\
P_f = 1 - \left(1 - P_e\right)^k. \\
\end{array} 
\right.
\label{eq:UnitaryUnscrambled}
\end{equation}

In order to estimate the effect of the non-systematic transmission induced by
the introduction of a scrambling matrix with row and column weight $> 1$, we
can first refer to an ideal case.
Such ideal case is what we call \textit{perfect scrambling}; it models
a scrambling technique that, in presence of one (or more) error(s), produces maximum uncertainty.
In other terms, under the hypothesis of perfect scrambling, a single residual
bit error in the decoded word is sufficient to ensure that half of its bits are
in error after descrambling.
In practice, perfect scrambling can be approached by using dense $\mathbf{S}$ (with dense $\mathbf{S}^{-1}$)
matrices, that is, with a high density of $1$ symbols.
A very high scrambling effect is obtained when the density of $\mathbf{S}^{-1}$
is $0.5$, but also a lower density could suffice to achieve an almost perfect
scrambling effect.

It is easy to prove that, under the condition of perfect scrambling, the bit
error rate after descrambling equals half the frame error rate
expressed by \eqref{eq:UnitaryUnscrambled}, that is:

\begin{equation}
P_e^{PS} = \frac{1}{2}\left\{1 - \left[1 - \frac{1}{2}\mathrm{erfc}\left(\sqrt{\frac{E_b}{N_0}}\right)\right]^k\right\}. \\
\label{eq:UnitaryPerfectScrambling}
\end{equation}

So, the condition of perfect scrambling can be used as a bound, since it
gives Eve's maximum bit error rate.
When we instead adopt a real scrambling matrix, the bit error rate
for a unitary rate coded transmission can be conveniently estimated.
First of all, it is necessary to evaluate the bit error rate conditioned to erred frames, since scrambling is effective only on them. Such probability can be expressed as:

\begin{equation}
P_r = \frac{P_e}{P_f}.
\end{equation}

If we denote by $w(i)$ the Hamming weight of the $i$-th column of $\mathbf{S}^{-1}$,
$w(i) \leq k, \forall i \in [1,\ldots,k]$, an approximate estimate of the bit error rate on
the $i$-th received bit after descrambling can be obtained by using arguments
similar to those developed in \cite{Gallager1963} and it is expressed:

\begin{equation}
P_e^{S}(i) = P_f\frac{1-{\left(1-2 P_r\right)}^{w\left(i\right)}}{2}.
\label{eq:UnitaryScrambling}
\end{equation}

\begin{figure}[t]
\begin{centering}
\includegraphics[width=83mm,keepaspectratio]{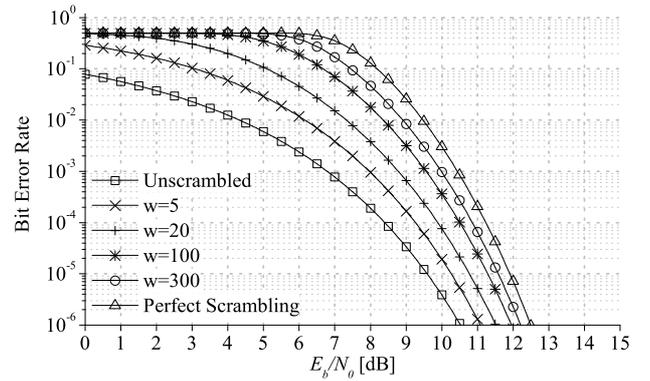}
\caption{Bit error rate with unitary rate coding ($k = n = 1576$) for several levels of scrambling. \label{fig:UnitaryRate}}
\par\end{centering}
\end{figure}

As a numerical example, we have considered the case $k = n = 1576$ (that will be of
interest in the following) and calculated the bit error rate for
several degrees of scrambling.
For the sake of simplicity, we have studied the case of regular scrambling
matrices, that is, $w(i) = w, \forall i \in [1,\ldots,k]$.
Fig. \ref{fig:UnitaryRate} shows the curves of $P_e^{S}$
for some values of $w$, and for the two limit cases of absence of scrambling
(unscrambled transmission) and perfect scrambling.

As we observe from the figure, the unscrambled transmission gives low values
of bit error rate even at rather low SNR.
On the contrary, by assuming a sufficiently large value of $w$, scrambling permits us to keep the bit error rate close to $0.5$ (corresponding to the case of complete lack of 
information) up to a rather high $\frac{E_b}{N_0}$ threshold. 
In addition, scrambling helps to improve the slope of the 
$P_e$ curve, so reducing the security gap. 
As expected, perfect scrambling ensures the best performance,
but an $\mathbf{S}^{-1}$ matrix with density $\approx 0.2$ ($w = 300$) is
sufficient to have a similar (optimal) behavior.

\subsection{\textit{t}-Error Correcting Coding}

In order to further improve the slope of the $P_e$ curves,
a linear block code with dimension $k < n$ can be introduced.

In this subsection, we consider the adoption of an $(n, k)$ linear block code
able to correct $t$ bit errors under hard-decision decoding.
Such code could be, for example, a Bose-Chaudhuri-Hocquenghem (BCH) code;
in the following we will consider the $(2047,1354)$ BCH code, able to
correct $t = 69$ errors. This code has a value of $k$ not so different from that considered in the previous subsection.

When such a coding scheme is adopted, the frame error rate and
bit error rate at the receiver can be estimated as follows:

\begin{equation}
\left\{ 
\begin{array}{l}
P_f = \displaystyle\sum^n_{i=t+1} {n \choose i} P_0^i (1-P_0)^{n-i}, \\ 
P_e = \displaystyle\sum^n_{i=t+1}\frac{i}{n} {n \choose i} P_0^i (1-P_0)^{n-i}, \\ 
\end{array} 
\right.
\label{eq:tErrorCoding}
\end{equation}
where $P_0$ is the channel bit error rate, taking into account the bandwidth
expansion due to the presence of the code:
\begin{equation}
P_0 = \frac{1}{2}\mathrm{erfc}\left(\sqrt{\frac{E_b}{N_0}\cdot\frac{k}{n}}\right). \\
\end{equation}

Starting from Eqs. \eqref{eq:tErrorCoding}, we can easily obtain that, in
presence of perfect scrambling, the bit error rate of a transmission
based on a $t$-error correcting code becomes:

\begin{equation}
P_e^{PS} = \frac{1}{2}P_f = \frac{1}{2} \displaystyle\sum^n_{i=t+1} {n \choose i} P_0^i (1-P_0)^{n-i}.
\label{eq:tErrorCodingPS}
\end{equation}

\begin{figure}[t]
\begin{centering}
\includegraphics[width=83mm,keepaspectratio]{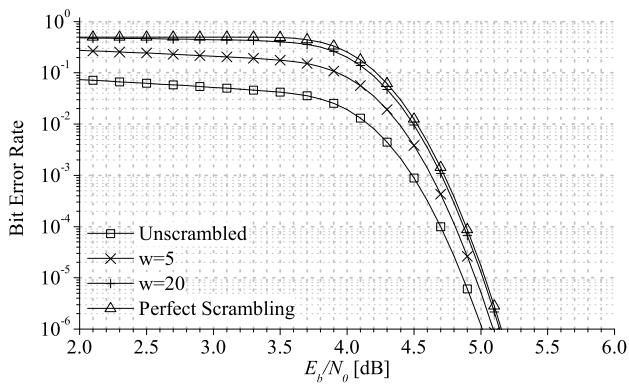}
\caption{Bit error rate for the $(2047,1354,69)$ BCH code and different levels of scrambling. \label{fig:tErrorCorrecting}}
\par\end{centering}
\end{figure}

Based on these modified expressions, Eq. (\ref{eq:UnitaryScrambling}) can be used again. Some examples are shown in Fig. \ref{fig:tErrorCorrecting} where we have considered a $(2047,1354,69)$ BCH code with different
levels of scrambling.
By a comparison with Fig. \ref{fig:UnitaryRate} (that, however, refers to a slightly different value of $k$) we see that the introduction of the code reduces the signal-to-noise ratio (as obvious and expected) and, mostly important for our purposes, increases the slope of the $P_e^{S}$ and $P_e^{PS}$ curves. Besides slope increase, scrambling contributes to emphasize the knee between the region of high bit error rate and that of low/medium bit error rate, that is the actual requirement for having a small secrecy gap. From Fig. \ref{fig:tErrorCorrecting} we see that $w = 20$ (density $w/k \approx 0.01$) is enough for this purpose, while in absence of the code (see Fig. \ref{fig:UnitaryRate}) $w = 300$ (density $w/k \approx 0.19$) was necessary.

\section{Non-Systematic LDPC codes}
\label{sec:four}

As an example of Soft-In Soft-Out modern error correcting schemes, we have considered
LDPC codes, to which we have applied the approach of non-systematic transmission based
on scrambling.
For the sake of comparison, we have also considered the approach based on puncturing
proposed in \cite{Klinc2009}.

Non-systematic LDPC codes have been already studied outside the
physical layer security issue.
In particular, they have been proved able to give an important
advantage over systematic encoding in the presence of source redundancy.
In \cite{Shamir2005, Alloum2005, Shamir2006, Shamir2006a} non-systematic LDPC codes
for redundant source data are studied.
Non-systematic encoding is accomplished by using the same scrambling approach considered
in this paper or through alternative techniques as \textit{post-coding} and \textit{splitting}.

A similar approach to the design of non-systematic LDPC codes is also presented in \cite{Lin2008},
where the authors aim at designing codes with sparse generator matrices, in
such a way that the bit error rate performance remains not far from that of systematic LDPC codes. We notice that such target is diametrically opposed to physical layer security.

In our simulations, we have considered two LDPC codes designed through the Progressive Edge
Growth (PEG) algorithm \cite{Hu2001PEG}.
Both codes have a lower triangular parity-check matrix, in such a way as to allow systematic
encoding without the need of Gaussian elimination.
The first code has length $n = 2364$, dimension $k = 1576$ (hence, code rate $R = 2/3$) and it 
has been used for simulation of systematic and non-systematic transmission. 
The latter has been obtained by adopting a dense $1576 \times 1576$ scrambling matrix, in order
to approach the effect of a perfect scrambler.
The second code has length $n = 3940$ and dimension $k = 1576$.
It has been used to simulate punctured transmission, by puncturing all its $1576$ information
bits. So, the transmission rate results in $1576 / \left( 3940 - 1576 \right) = 2/3$, as for
the cases without puncturing.
\begin{figure}[t]
\begin{centering}
\includegraphics[width=83mm,keepaspectratio]{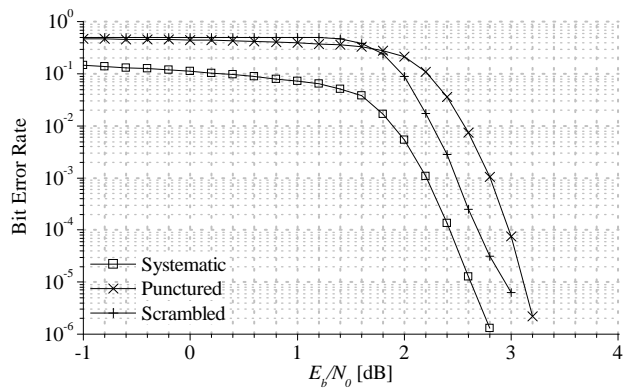}
\caption{Simulated bit error rate for LDPC codes with $k = 1576$ and $R = 2/3$, adopting systematic transmission, puncturing and scrambling. \label{fig:LDPC}}
\par\end{centering}
\end{figure}

Fig. \ref{fig:LDPC} shows the simulated performance, in terms of bit error rate, for the
considered transmission schemes based on LDPC codes.
As we observe from the figure, the systematic transmission ensures the best performance in
terms of error correction capability. However, in the considered context of physical layer
security, it shows an important drawback, that is, a bit error rate significantly smaller than $0.5$ even at low 
signal-to-noise ratio (the same behavior was observed in Figs. \ref{fig:UnitaryRate} and \ref{fig:tErrorCorrecting}).

The approach based on puncturing gives worse error correcting performance,
with a loss of about $0.5$ dB in the waterfall region with respect to systematic
LDPC coding.
However, the usage of punctured bits for transmitting the secret message is able to
ensure a higher bit error rate for low signal-to-noise ratio.
Both such aspects could be improved by adopting non-systematic unpunctured LDPC codes
based on the proposed scrambling technique: the performance loss with respect
to the systematic LDPC code is about $0.3$ dB in the waterfall region, and
the bit error rate is maintained close to $0.5$ in a larger interval of
signal-to-noise ratio values.

These facts reflect on the security gap over the AWGN wire-tap channel.
Discussion of the security gap and comparison of the considered techniques from this point of view
are reported in the next section.

\section{Comparison between the considered techniques}
\label{sec:five}

\begin{figure}[t]
\begin{centering}
\includegraphics[width=83mm,keepaspectratio]{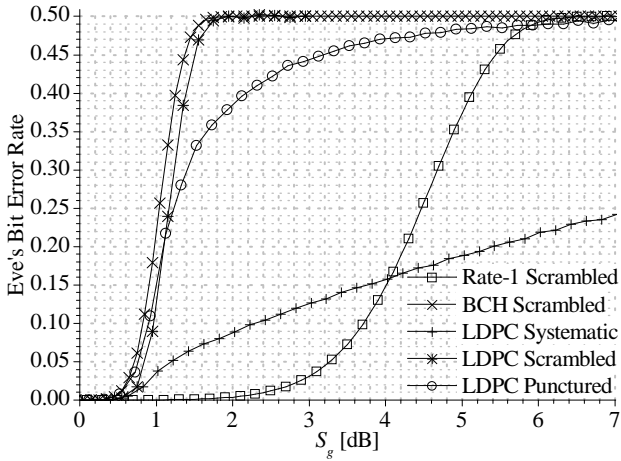}
\caption{Bit Error Rate versus the secrecy gap for the techniques considered. \label{fig:BERvsGap}}
\par\end{centering}
\end{figure}

In order to compare the considered transmission techniques, we have fixed
$\overline{P_e^\mathrm{B}} = 10^{-5}$, and calculated $\left. \overline{\frac{E_b}{N_0}} \right|_\mathrm{B}$.
Starting from the value of $\left. \overline{\frac{E_b}{N_0}} \right|_\mathrm{B}$, it
is possible to estimate Eve's bit error rate $P_e^\mathrm{E}$ as
a function of the gap $S_g$.
Fig. \ref{fig:BERvsGap} reports these curves for the considered transmission techniques.
In the figure, for all techniques that adopt scrambling, the perfect scrambling condition
has been considered (simulation of the scrambled LDPC code has been done with a dense
$\mathbf{S}^{-1}$ matrix, able to approach perfect scrambling).
As we observe from the figure, the usage of a systematic LDPC code gives a very slow
convergence of Eve's bit error rate to the ideal value of $0.5$.
So, such technique requires a very high security gap for realistic values of $\overline{P_e^\mathrm{E}}$
(that are usually $\geq 0.4$).
The reason of such a slow convergence is systematic transmission: if we adopt
a non-systematic unitary rate code, even renouncing any error correction capability,
performance is improved and $P_e^\mathrm{E} \geq 0.4$ is reached for a gap value around $5$ dB.
The situation can be further improved by adopting non-systematic error correcting codes.
If we implement non-systematicity through puncturing, the condition $P_e^\mathrm{E} \geq 0.4$
is achieved at a $2.2$ dB gap.

The best performance is achieved by implementing non-systematic coded transmission through
scrambling. Both the BCH and the LDPC code, under the condition of perfect scrambling,
give very good performance.
The condition $P_e^\mathrm{E} \geq 0.4$ is reached at $1.3$ dB and $1.4$ dB gap by the
scrambled BCH and LDPC code, respectively. Obviously, LDPC codes have the advantage of permitting us to work at smaller SNR.

\section{Conclusion}
\label{sec:six}
We have investigated the usage of non-systematic codes for achieving physical layer security.
We have focused on the AWGN wire-tap channel model, and estimated the security gap
as a measure of the effectiveness of several transmission schemes.

Our results show that systematic coded transmission (as well as uncoded transmission) is
unsuited to such kind of applications, due to the low bit error rate values it
achieves even at low signal-to-noise ratio.

Non-systematic transmission, instead, is able to reduce the security gap in terms of
signal-to-noise ratio that is needed between Bob's and Eve's AWGN channels in order to
achieve physical layer security.
We have compared non-systematic transmission implemented through scrambling and puncturing,
and showed that the former is able to outperform the latter, requiring a smaller security
gap.

\newcommand{\BIBdecl}{\setlength{\itemsep}{0.01\baselineskip}}
\bibliographystyle{IEEEtran}

\begin{thebibliography}{10}
\providecommand{\url}[1]{#1}
\csname url@samestyle\endcsname
\providecommand{\newblock}{\relax}
\providecommand{\bibinfo}[2]{#2}
\providecommand{\BIBentrySTDinterwordspacing}{\spaceskip=0pt\relax}
\providecommand{\BIBentryALTinterwordstretchfactor}{4}
\providecommand{\BIBentryALTinterwordspacing}{\spaceskip=\fontdimen2\font plus
\BIBentryALTinterwordstretchfactor\fontdimen3\font minus
  \fontdimen4\font\relax}
\providecommand{\BIBforeignlanguage}[2]{{%
\expandafter\ifx\csname l@#1\endcsname\relax
\typeout{** WARNING: IEEEtran.bst: No hyphenation pattern has been}%
\typeout{** loaded for the language `#1'. Using the pattern for}%
\typeout{** the default language instead.}%
\else
\language=\csname l@#1\endcsname
\fi
#2}}
\providecommand{\BIBdecl}{\relax}
\BIBdecl

\bibitem{Wyner1975}
A.~D. Wyner, ``The wire-tap channel,'' \emph{Bell Syst. Tech. J.}, vol.~54,
  no.~8, pp. 1355--1387, Oct. 1975.

\bibitem{Maurer1993}
U.~Maurer, ``Secret key agreement by public discussion from common
  information,'' \emph{{IEEE} Trans. Inform. Theory}, vol.~39, no.~3, pp.
  733--742, May 1993.

\bibitem{Thangaraj2007}
A.~Thangaraj, S.~Dihidar, A.~Calderbank, S.~McLaughlin, and J.-M. Merolla,
  ``Applications of {LDPC} codes to the wiretap channel,'' \emph{{IEEE} Trans.
  Inform. Theory}, vol.~53, no.~8, pp. 2933--2945, Aug. 2007.

\bibitem{Merhav2008}
N.~Merhav, ``{Shannon}'s secrecy system with informed receivers and its
  application to systematic coding for wiretapped channels,'' \emph{{IEEE}
  Trans. Inform. Theory}, vol.~54, no.~6, pp. 2723--2734, Jun. 2008.

\bibitem{Klinc2009}
D.~Klinc, J.~Ha, S.~McLaughlin, J.~Barros, and B.-J. Kwak, ``{LDPC} codes for
  physical layer security,'' in \emph{Proc. {IEEE} Global Telecommunications
  Conference (GLOBECOM 2009)}, Honolulu, HI, Nov. 2009, pp. 1--6.

\bibitem{Kwak2009}
B.-J. Kwak, N.-O. Song, B.~Park, D.~Klinc, and S.~McLaughlin, ``Physical layer
  security with {Yarg} code,'' in \emph{Proc. First International Conference on
  Emerging Network Intelligence}, Sliema, Malta, Oct. 2009, pp. 43--48.

\bibitem{Harrison2009}
W.~Harrison and S.~McLaughlin, ``Physical-layer security: Combining error
  control coding and cryptography,'' in \emph{Proc. {IEEE} International
  Conference on Communications (ICC '09)}, Dresden, Germany, Jun. 2009, pp.
  1--5.

\bibitem{Leung-Yan-Cheong1978}
S.~Leung-Yan-Cheong and M.~Hellman, ``The {Gaussian} wire-tap channel,''
  \emph{{IEEE} Trans. Inform. Theory}, vol.~24, no.~4, pp. 451--456, Jul. 1978.

\bibitem{Lai2008}
L.~Lai, H.~El~Gamal, and H.~V. Poor, ``The wiretap channel with feedback:
  Encryption over the channel,'' \emph{{IEEE} Trans. Inform. Theory}, vol.~54,
  pp. 5059--5067, Nov. 2008.

\bibitem{McEliece1978}
R.~J. McEliece, ``A public-key cryptosystem based on algebraic coding theory.''
  \emph{DSN Progress Report}, pp. 114--116, 1978.

\bibitem{Gallager1963}
R.~G. Gallager, \emph{Low Density Parity Check Codes}.\hskip 1em plus 0.5em
  minus 0.4em\relax Cambridge, MA: M.I.T. Press, 1963.

\bibitem{Shamir2005}
G.~Shamir and J.~Boutros, ``Non-systematic low-density parity-check codes for
  nonuniform sources,'' in \emph{Proc. International Symposium on Information
  Theory ({ISIT 2005})}, Adelaide, Australia, Sep. 2005, pp. 1898--1902.

\bibitem{Alloum2005}
A.~Alloum, J.~Boutros, G.~Shamir, and L.~Wang, ``Non-systematic {LDPC} codes
  via scrambling and splitting,'' in \emph{Proc. {Allerton}'s Conference on
  Communication and Control}, Monticello, Illinois, Sep. 2005, pp. 1879--1888.

\bibitem{Shamir2006}
G.~Shamir, J.~Boutros, A.~Alloum, and L.~Wang, ``Non-systematic {LDPC} codes
  for redundant data,'' in \emph{Proc. Inaugural Workshop for the Center of
  Information Theory and its Applications}, San Diego, California, Feb. 2006.

\bibitem{Shamir2006a}
G.~Shamir, L.~Wang, and J.~Boutros, ``High rate non-systematic {LDPC} codes for
  nonuniform sources,'' in \emph{Proc. 4th International Symposium on Turbo
  Codes and Related Topics}, Munich, Germany, Apr. 2006.

\bibitem{Lin2008}
D.~Lin, Q.~Li, and S.~Li, ``Construction of nonsystematic low-density
  parity-check codes based on symmetric balanced incomplete block design,''
  \emph{Journal of Electronics (China)}, vol.~25, no.~4, pp. 445--449, Jul.
  2008.

\bibitem{Hu2001PEG}
X.~Y. Hu and E.~Eleftheriou, ``Progressive edge-growth tanner graphs,'' in
  \emph{Proc. {IEEE} Global Telecommunications Conference ({GLOBECOM}'01)}, San
  Antonio, Texas, Nov. 2001, pp. 995--1001.

\end{thebibliography}

\end{document}